# Potentials of Electric Vehicles for the Provision of Active and Reactive Power Flexibilities as Ancillary Services at Vertical Power System Interconnections

Manuel WINGENFELDER*, Marcel SARSTEDT, Lutz HOFMANN
Leibniz University Hannover,
Institute of Electric Power Systems, Power Engineering Section
Appelstr. 9A, 30167 Hanover, Germany
[surname]@ifes.uni-hannover.de

## SUMMARY

Active and reactive power flexibilities of decentral energy resources or controllable loads connected to the low and medium voltage level can be used to provide ancillary services for the higher level system operator at the vertical system interface. Therefore, the lower level grid operator aggregates the flexibility potentials of these distributed flexibility providing units (FPU) and their influence on the vertical active and reactive interconnection power flows as a polygon within the PQ-plane under consideration of technical constraints. This so called feasible operation region (FOR) is used by the higher level grid operator for the specification of a flexibility demand and ancillary service provision at the vertical system interconnection in the context of system operator cooperation.

This paper extends the research regarding the determination of the feasible operation region under the impact of electric vehicles (EV). Thereby, the active and reactive power flexibility potentials of EV for alternating current and direct current bidirectional charging are limited in first instance by regulatory guidelines. In this paper German regulatory guidelines are applied to depicture the current state of the art for EV charging components. Under assumption of bidirectional charging for EV, the corresponding flexibility polygon in the PQ-plane is derived. The impact on the FOR is then evaluated within the Cigré European medium voltage benchmark system. In a first step, the flexibility potential of the system is determined for a load case as basis for the comparison with different, realistic penetration levels of EV. DER are considered to depict additional existing power flexibilities. In a second step, the FOR and the corresponding limitations within the system are analyzed. Under variation of the rated line current, the limiting utility constraints are pointed out. The obtained results are the basis for more detailed investigations regarding the potentials of bidirectional charging for an intensified system operator's cooperation. The used method presents a promising approach to be used by the lower system operator as tool in the grid planning process to estimate the possible power flexibilities that can be provided to a higher system.

Continuing, the potential for ancillary services provided by EV is discussed, resulting in the derivation of future research aspects. Because the accumulation of EV leads to higher grid utilization, which results in quadratically increasing grid losses the closer the grid is operating to the grid constraints, a cost curve for the EV's active power flexibility is introduced, whereas probability zones are indicated.

The main contribution of this paper is to merge the possibility of using the power flexibility of EV with an existing method for the determination of the FOR. This concludes in aspects that present subjects of future research, which are generally approachable with the applied tools.

# 1 INTRODUCTION

Decentral energy resources (DER) provide ancillary services for local voltage maintenance and reactive power management within an active distribution grid level. The distributed flexibility provision of the flexible providing units (FPU) influences the active and reactive interconnection power flows, which can be used for an ancillary services provision across several voltage levels in the context of intensified system operator cooperation. A cooperation between the distribution system operators (DSO) and transmission system operator (TSO) is fundamentally enabled by the aggregation of the flexibility potentials of FPU within the distribution system level by a so called feasible operation region (FOR) in the PQ-plane (see **Figure 1**, cf. [1], [2]). Generally, the FOR is limited either due to the maximum amount of provided power flexibility by the existing FPU or due to the grid operation limits regarding voltage constraints and current limits of lines and transformers. Flexible loads and DER, such as wind turbines or photovoltaic units (PV), offer certain degrees of flexibility in dependency of their placement. FPU with a certain rated power are located in the medium voltage (MV) level and are aggregated as a FOR to be offered to the higher system operator. These flexibilities can be requested by the TSO for its system management. After a specification of an operation point, the requested vertical active power $P_{\text{vert}}$ and reactive power $Q_{\text{vert}}$ are distributed over the available FPU. Smaller rated FPU are connected at LV level, which are preliminary aggregated within the lower system level and are also integrated in the process. The queuing of the FPU follows a bidding process. Market design recommendations for ancillary services are already introduced within certain options in [3]. The provision of local ancillary services offers a reduced utility bill for customers on the one hand and, on the other hand, has the potential to delay costly grid reinforcement.

The share of electric vehicles (EV) is on the rise worldwide [4] to reduce $CO_2$ emissions within the mobility sector [5]. Consequently, the roll-out of charging infrastructure has risen. E.g., in Germany, boosted by governmental funding, the amount of publicly accessible charging infrastructure multiplied between 2016 and the beginning of 2021 by 1,268% concluding in 42,388 charging points [6]. The charging demand of EV challenges system operators to provide a sufficient active power supply while operating the electrical grid without violating the grid constraints. Flexible charging offers a high potential regarding an optimized integration into the distribution grid level. While flexible unidirectional charging focuses on a reduced simultaneity and offers the application of scheduled charging concepts, flexible bidirectional charging enables further applications. Generally, it has been reviewed in [7] and shown for voltage regulation in [8]. Flexible charging demand has been evaluated and emphasized in multiple investigations [9], [10], [11] to either be considered in congestion management or optimized usage of generated power provided by decentral energy resources [12]. Due to controlled charging, electrical vehicles function at the lower distribution grid levels as FPU, which add to the FOR in the context of the hierarchical multi-level grid control strategies. The influence of power flexibility polygon shapes of new FPU on an existing FOR in general has been evaluated in [13]. Thus, this paper extends the research regarding the determination of the FOR under the impact of EV within a case study by investigating different scenarios for the EV penetration under consideration of preexisting flexible loads and available DER.

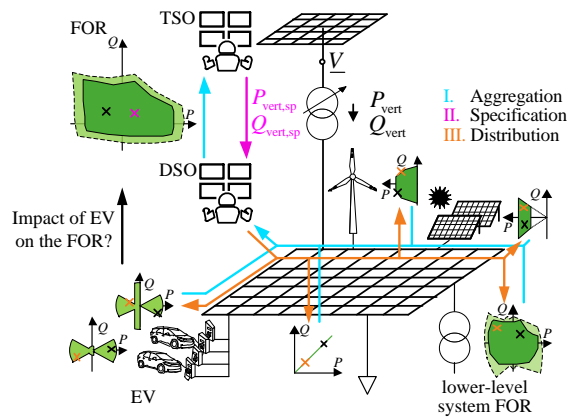

**Figure 1**: Schematic system interconnection for DSO and TSO interface with exemplary extended FOR by EV (adapted from [1])

## 2 TECHNICAL AND REGULATORY FLEXIBILITY OF EV CHARGING

The alternating current (AC) and direct current (DC) charging power of EV is limited by the rated power of the inverter, resulting in a circle or for higher rated components in a square in the PQ-plane. A limitation of the available reactive power provision is given by the effective connecting inductance [14]. DC charging infrastructure uses inverters, which are located external of the EV. Certain types can be used in STATCOM operation, which offers an extensively decentralized grid support, when distributed in analogy to local load demand. However, AC charging is limited by the built-in inverter in the EV, which is assumed to be realized by using minimal technical requirements.

While the technical potential is often higher, the utilizable potential is limited by technical regulations, which change in respect to country and time. As it provides a suitable example and will also be mandatory under applied standard, German VDE application guidelines [15], [16] are used in this case study to consider regulatory guidelines for the power flexibilities for AC charging (**Figure 2**) and DC charging (**Figure 3**) as well as for feed-in, when connected to the distribution grid. While there is no specification for the reactive power $Q$ during low active power infeed $P < 0.2\,P_n$ (for $P < 0$) and withdrawal $P < 0.05\,P_n$ (for $P > 0$), the VDE application guidelines limit the feasible reactive power flexibility for active power above the given values by a $\cos(\varphi)$ of 0.9 (ind. and cap.) at rated powers of charging stations over 13.8 kVA. For a rated power lower or equal to 13.8 kVA, a $\cos(\varphi)$ of 0.95 or better should be maintained when discharging.

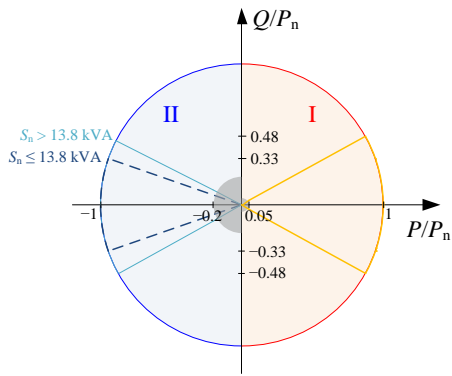

**Figure 2**: Active and reactive power flexibility for AC charging under consideration of German application guidelines

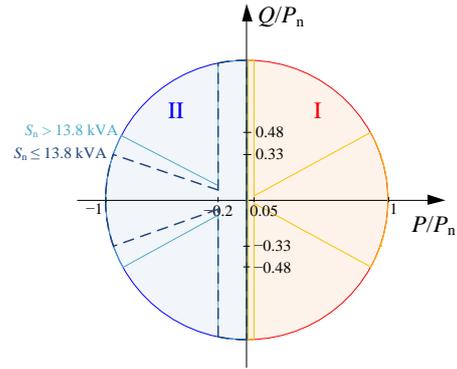

**Figure 3**: Active and reactive power flexibility for DC charging under consideration of German application guidelines

## 3 CASE STUDY

The basis for the case study is an adapted version of the Cigré European medium voltage benchmark test system [17]. To represent a preliminary active and reactive power flexibility in the PQ-plane, loads and DER have been considered as flexibility providing units in addition to the conventional test system. The locations, the flexibilities and the operating points of the FPU, e.g. residential loads, industrial loads, wind turbines and photovoltaic units, are depicted in **Figure 4**. Compensation and storage units are neglected. Active power values are given in MW and reactive power values in Mvar. The high voltage bus of the transformer is the slack node with a nominal voltage $\underline{V}_{\text{slack}}$ of 110 kV $e^{j0}$. Grid operation limits in regards of voltage constraints are set to 0.9 and 1.1 in pu concerning the nominal voltage. Data on lines and transformers are given in **Table I** and **Table II**, respectively. The full data set is accessible under [18]. For the determination of the FOR various approaches were introduced in the recent years. A detailed overview is given in [2] and [19]. In this paper, an existing simulation environment (cf. [2]) with a particle swarm optimization based aggregation method and various sampling strategies is used for the determination of the FOR. Besides an appropriate quality of the results within acceptable computation time, the main argument to choose this metaheuristic approach is the possible consideration of non-convex FPU flexibility polygons (cf. [20]) as provided by EV (see chapter 2).

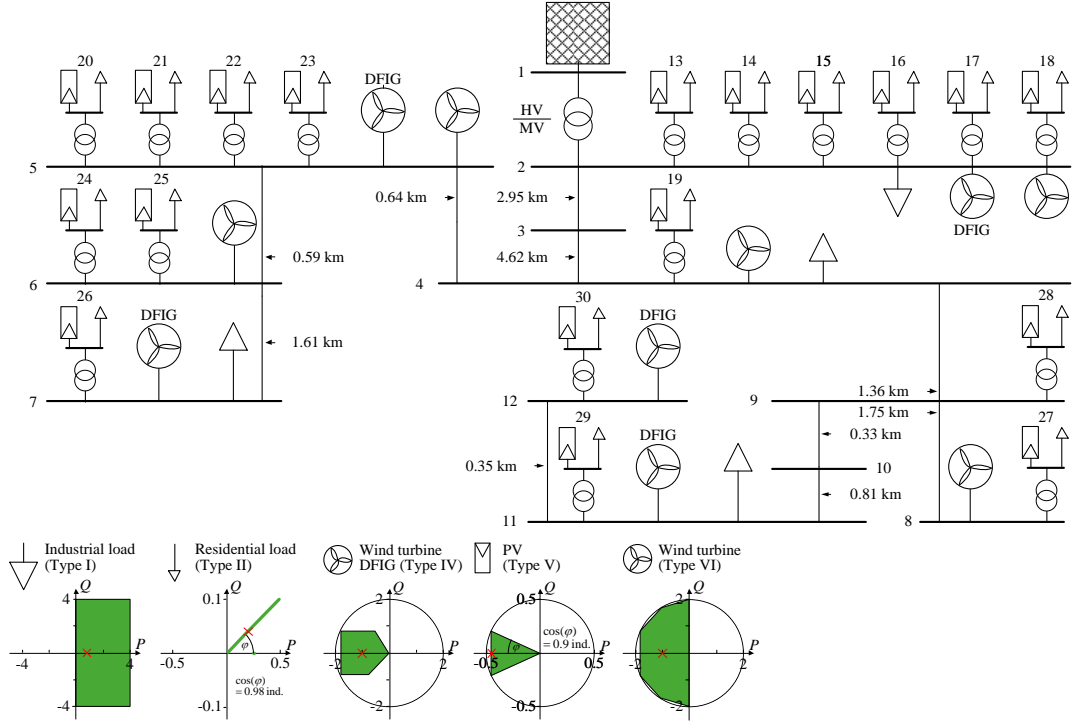

**Figure 4**: Adapted topology of Cigré European MV distribution network benchmark system including power flexibilities and operating points of the FPU

**Table I**: Line data

| Length specific line resistance $R'$ | Length specific line inductance $L'$ | Length specific line capacitance $C'$ |
|---|---|---|
| 0.501 Ω/km | 2.279 mH/km | 0.151 μF/km |

**Table II**: Transformer data

| Nominal upper voltage $V_{HV}$ | Nominal lower voltage $V_{LV}$ | Rated apparent power $S_r$ | Short-circuit voltage $v_{sc}$ | Copper losses $P_{Cu}$ | Open-circuit current $i_{oc}$ | Iron core losses $P_{Fe}$ |
|---|---|---|---|---|---|---|
| 110 kV | 20 kV | 25 MVA | 12% | 25 kW | 0.5% | 0 kW |
| 20 kV | 0.4 kV | 2 MVA | 8% | 16.7 kW | 0.2% | 4 kW |

### A. SCENARIOS FOR FPU

As a basic scenario, the industrial and residential loads with their power flexibility and operating point as shown in **Figure 4** are connected constantly to the grid. The power flexibilities provided by the DER (wind turbines and PV) as well as their operating points are either set as depicted or set to zero to represent their availability due to weather conditions.

EV shares are developed based on the load data set. A recalculation for the number of considered households is done with simultaneity factors [21] to derive considerable numbers. A considerably high simultaneity factor $g_\infty$ of 0.7 and a peak active power value $P_p$ of 15 kW are chosen to represent modern households. Calculating with 0.5 MW as the maximum load at low voltage nodes, the corresponding amount of households is derived to 50. It is assumed that in average each household holds one car, which represents roughly the private vehicle share in regards of households in Germany [22]. The shares of EV to the whole number of passenger cars are set to 10%, 20% and 30%. It is applied that AC charging with a rated apparent power of 11 kVA is available for each EV at the LV nodes. For MV nodes, certainly where small industrial loads are aggregated, EV loads are assumed as DC charging infrastructure with 600 kVA each. The operating point for both charging types is set to the maximum rated power with a $\cos(\varphi)$ of 0.999 to represent charging behavior of modern EV. The concluding numbers for EV with respect to the nodes are presented in **Table III**.

Overall, the considered scenarios for DER and EV in regards of the FOR shapes in the following section are shown in **Table IV**.

Table III: Considered EV cases for AC charging at LV nodes and DC charging at MV nodes

| Case Nr. | AC charging $S_r$ in kVA | LV nodes | DC charging $S_r$ in kVA | MV nodes |
|---|---|---|---|---|
| 1 | 55 | all | 600 | 2 |
| 2 | 110 | all | 600 | 2, 4 |
| 3 | 165 | all | 600 | 2, 4, 7 |

Table IV: Scenarios for FPU

|  | No DEA available | All DEA available |
|---|---|---|
| No EV | scenario 0 | scenario 0a |
| EV Case 1 | scenario 1 | scenario 1a |
| EV Case 2 | scenario 2 | scenario 2a |
| EV Case 3 | scenario 3 | scenario 3a |

### B. SCENARIOS FOR RATED LINE CURRENT

The maximum permitted continuous line current $I_{th}$ is varied to express the influence of changing grid utility constraints on the FOR, while FPU are available. Thus, the rated line current is fundamentally set to 220 A and manipulated for further results to 680 A.

## 4 RESULTS

The resulting FOR at the vertical DSO/TSO-system interconnection varies within the previously defined scenarios. Firstly, the results are presented regarding the scenario variations for EV and DER to conclude on the impact different FPU have on the FOR. Secondly, the constraints of the grid utilities are analyzed to better understand the results and to show a practical use-case of this method. Each operation point in the FOR is feasible and offers a secured flexibility potential.

### A. INFLUENCE OF FPU

In the following, the influence of adding FPU to preexisting flexible loads is analyzed. The FOR of zone I in **Figure 5** originates from the flexibilities of industrial and residential loads (cf. **Figure 4**) representing as scenario 0 the basis for the comparison. When adding EV according to scenario 3, zone II emerges. The FOR is limited either by the maximum provided active power flexibility (for $P < 0$) or the line current constraints (for $P > 0$). The upper and lower voltage limits can be expanded by the reactive power flexibility provided by the integrated EV. When applying scenario 0a by adding DEA to the initial state of zone I, it is expanded to zone III. In this case, the provided flexibility of all FPU in either $P$ direction is fully utilized without being limited by line current constraints. Grid operation voltage limits are restricting the FOR in either $Q$ direction. Finally, under consideration of EV and DER as for scenario 3a the FOR results in zone IV and is limited fully by grid constraints. The provided reactive power flexibility of the EV expands zone III in all directions.

The integration of EV as FPU can be assumed as a follow-up process after integrating DER to the distribution grid. In that case, the DER provide power flexibilities under fitting weather conditions in addition to flexible loads (scenario 0a), resulting in the FOR in **Figure 6** (zone III). It is expanded by integrating EV to the grid according to the defined scenarios 1a, 2a and 3a. Concerning the voltage constraints, the additional reactive power flexibility offers an increase in either $Q$ direction. In regards of the grid utility current constraints, it is notable in either $P$ direction, that the line current constraints are reached with increasing EV numbers. In total, the FOR is expanded by 11.49% due to the considered amount of EV in scenario 1a compared to only DER and loads of scenario 0a. With increasing EV, it grows by another 3.51% in scenario 2a and by additional 3.19% in scenario 3a. The FOR eventually reaches a maximized area, that can only be furtherly increased by grid reinforcement or expansion.

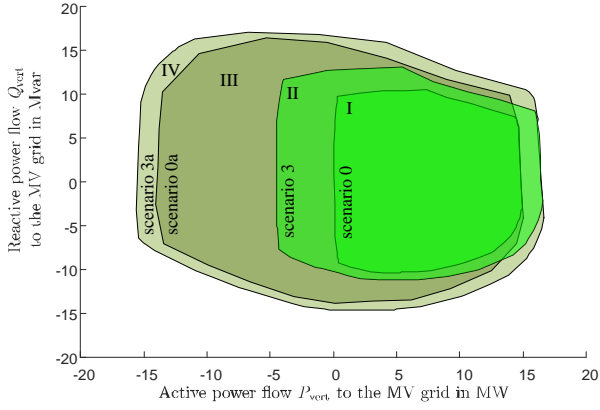

**Figure 5**: Simulation results of the FOR while considering the flexibility of loads (I); loads and EV (II); loads and DER (III); loads, EV and DER (IV) with $I_{th}$ = 220 A.

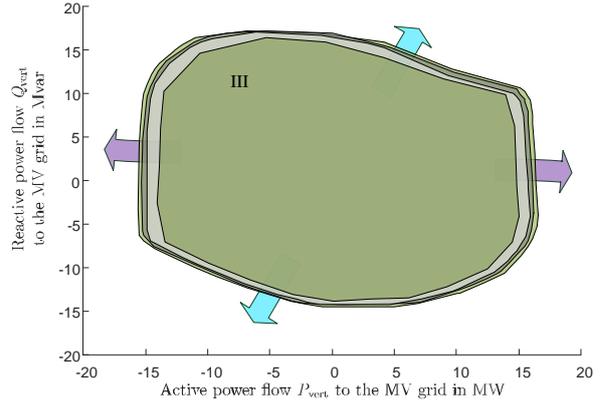

**Figure 6**: Simulation results of the FOR under consideration of all loads and all DER with increasing EV numbers over scenarios 0a, 1a, 2a and 3a

### B. INFLUENCE OF RATED LINE CURRENT

Considering all loads, DER and EV (scenario 3a) to exhaust the grid limits, the FOR results in **Figure 7** while the maximum permitted continuous line current is $I_{th}$ = 220 A (cf. zone IV, **Figure 5**). It is bounded for $Q < 0$ by the allowed maximum voltage (a)) and for $Q > 0$ by the allowed minimum voltage (b)). In either P direction, the FOR is bounded by the rated current of the transformer (c)) or by the rated line currents (d)). In this case, it is solely limited by $I_{th}$. When manipulating $I_{th}$ to 680 A, the FOR changes as shown in **Figure 8**. Now, the FOR is limited only in certain cases by the maximum permitted line current (d)), but mostly the transformer rating is the limiting factor (c)). Therefore, the deviation in flexibility by varying grid parameters can be mapped. By using the FOR, the lower system operator can easily identify the possible increase in flexibility by reinforcing specific grid equipment until the next technical condition is met. Also, possible flexibilities as valuable grid information is shared without directly sharing sensitive grid data such as topology or detailed utility parameters, which is crucial in the future collaboration between system operators.

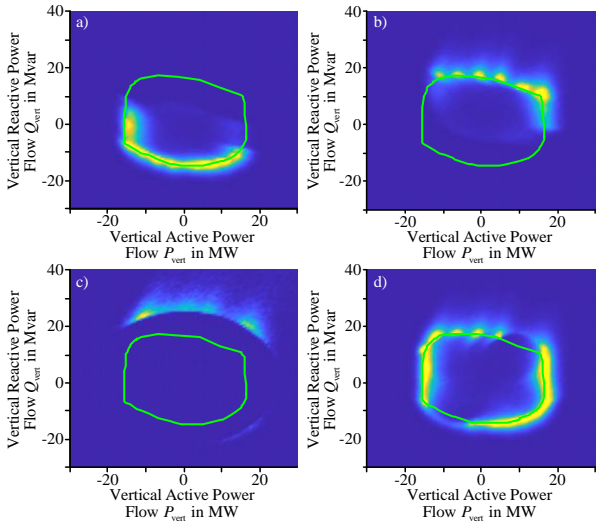

**Figure 7**: Simulation results of the FOR with loads, DER and EV (scenario 3a) and $I_{th}$ = 220 A, separated by constraints

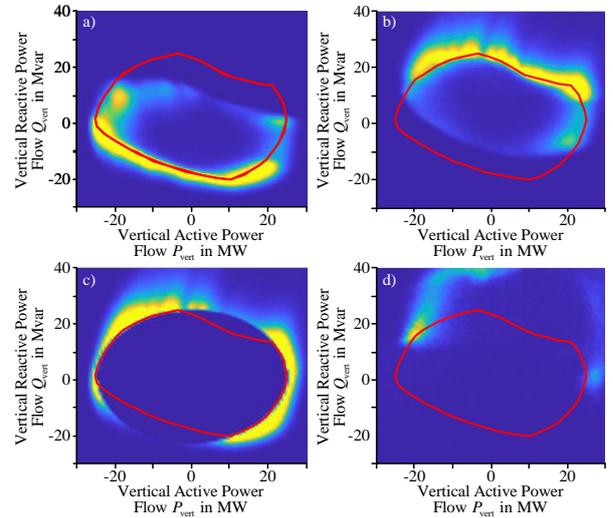

**Figure 8**: Simulation results of the FOR with loads, DER and EV (scenario 3a) and $I_{th}$ = 680 A, separated by constraints

## 5  DISCUSSION

The introduced approach for the active and reactive power flexibility provided by EV is assumed as a state of the art concept, so that it could be applied for today's EV and charging infrastructure. Manufacturers of these are not obligated to comply with regulatory guidelines, but it is assumed as aspired as it is in their interest to have reasonable sales figures, which would naturally decline, if the grid connection is refused. Also, the used guideline considerations are expected to be defined by regulatory instances in the near future.

The observed variations within the results of power flexibilities are as expected when applying the suggested scenarios. The results show that EV with flexible charging have a considerably impact on the FOR. Their influence is also notable when multiple FPU already offer a high degree of power flexibility. Especially in either reactive power direction, the existing voltage operation constraints are repositioned, due to the decentral character of the integrated EV. From these results the potential of only unidirectional charging can be derived as follows. The FOR would only be expanded in either $P$ direction while maintaining the enlarged $Q$ flexibility. Furthermore, the non-convexity behavior at the lower voltage constraint is notable, which has to be considered in the determination of the FOR (see e.g. [23]).

Different occupancies of charging stations offer dissimilar degrees of power flexibility. In particular, the controllability of charging in private charging infrastructure needs to meet with the end users demand. When mobility is wanted or needed, an available state of charge for the vehicle's battery has to be provided. Also, EV are expected to be mostly connected during night times at private charging infrastructure. Charging infrastructure installed at the work place for employee charging or fleet charging of the business EV is suggested to offer a higher flexibility within limited periods of the day or during the night. Public charging infrastructure is mostly used during the time of charging. Thus, the potential power flexibility is minor to not existent, since it would contradict to the user's demand. The quantification of power flexibility provided by EV for the various types of use has to be furtherly regarded in specific grid studies under scenario analysis.

Varying the rated line current represents grid reinforcement. This results in changed parameters for the lines, e.g. the resistivity and inductivity, which influences the FOR furtherly in regards of the voltage limits. The effect was neglected in this paper to show the usability of the FOR as a grid planning tool more strikingly. The FOR in specific and real grid topologies can be determined by the lower level system operator to participate in the flexibility market. This may become more feasible in the future, when distribution grids are working at their limits more commonly caused by the highly fluctuant loads, DER and EV. E.g. uncontrolled charging results in higher active power losses [9], [24]. The active power loss of the case study under consideration of loads, DER and EV according to scenario 3a is shown in **Figure 9** to underline the importance of realizing flexibilities in the near future as the losses increase quadratically from the center to the edges. For this instance, the participation of FPU in the market are based upon their cost factors, which is a crucial aspect to use the flexibility potential of EV as ancillary services within the power system. Therefore, the cost factor for EV is described in the following section as a deduction of further aspects.

## 6  ASPECTS OF MONETARIZING THE EV FLEXIBILITY

The monetarization of aggregated power flexibilities for loads and DER as well as for reactive power compensation and storages is examined in [20]. Hence, the monetarization of service costs and the expected payment function (EPF) for EV need to be developed.

In [25] different pricing models are discussed. For industrial customers, price models are incentive based approaches with direct control of large loads extended by market-orientated demand response (DR) with participation in demand bidding activities. Small customers (commercial and residential sector) more likely participate in price based programs. The dynamic pricing for the DR of charging demand (with focus on local generation of PV) based on deadline differentiated pricing is applied in [26], where values for the cost factor $c_P$ were set to 35 ct/kWh with either normal distribution or uniform distribution. The price is bound to the electricity market prices on low-voltage customer level until it gets decoupled from the existing billing system. Compared to a typical German market price for losses of 50 €/MWh, the option of EV's active power flexibility becomes quite expensive. Nevertheless, different price structures or applications using optimization methods may result in reasonable business models [27].

**Figure 10** shows three proposals for the EPF. Different EPF can lead to specialized pricing models, therefore a linear curve (Eq. 1), a piece-wise quadratic curve (Eq. 2) and a cubic curve (Eq. 3) are introduced, where $C_i$ is the total costs, $c_P$ the cost factor and $P$ the active power flexibility.

$$C_{\text{lin}} = c_{\text{P}} \cdot (1 - P) \tag{1}$$

$$C_{\text{quad}} = \begin{cases} c_{\text{P}} \cdot (P^2 + 1), & P \in [-1, 0[ \\ c_{\text{P}} \cdot (P - 1)^2, & P \in [0, 1] \end{cases} \tag{2}$$

$$C_{\text{cub}} = c_{\text{P}} \cdot (1 - P^3) \tag{3}$$

A deviating occurrence of active power flexibility demand of the EV in terms of grid-serving is indicated by different zones in **Figure 10**. Active power feed-in by the EV will only be required, if the grid is stressed load-wise and other feed-in sources are unavailable. Existing distribution grids are expected to be designed for the occurring load peaks, so the height of active power feed-in would be assumingly remote. If new loads are integrated before grid reinforcement is applied, the height and probability of active power feed-in rises. For active power draw-off there are two possibilities that would demand the EV flexibility. Either the feed-in power from DEA shall be increased for better usage of renewable energy (charging activates) or when the grid is already working at the current or voltage constraints and an additional unflexible load is connected (charging power decreases). Assumingly, the zones i and v are demanded the least often, the zones ii, iv and vi are demanded more commonly and the zones iii and vii are demanded the most often.

The cost factor for reactive power flexibility $c_Q$ for wind farms and its independent EPF have been evaluated [28], [29]. When neglecting technical specific reasons $c_Q$ can be assumed as approximately 1/100 of $c_P$ (cf. [20]). The value for EV will orientate on future bidding system, where DER will participate.

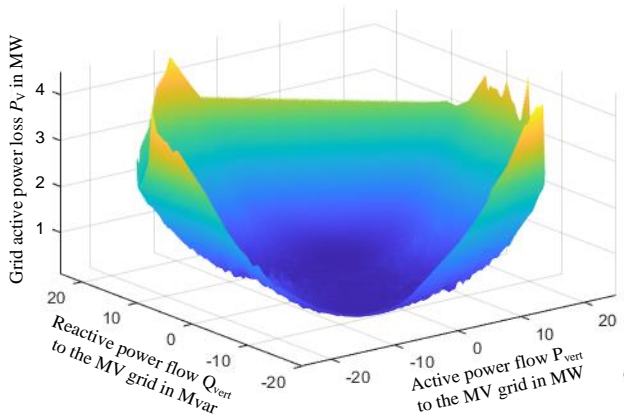
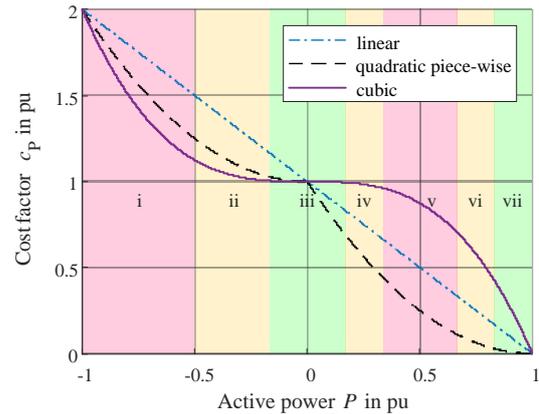

**Figure 9**: Active power loss of scenario with loads, DER and EV according to scenario 3a

**Figure 10**: Expected payment functions for EV power flexibility and expectable occurrence zones

## 7 CONCLUSION

Flexibility potentials of flexibility providing units (FPU) are aggregated by the lower level system operator at the vertical system interface as a feasible operation region (FOR), resulting in a polygon within the PQ-plane under consideration of technical constraints. Under application of concepts for AC and DC bidirectional charging from technical guidelines and realistic penetration scenarios, the impact of flexible bidirectional charging for EV on the FOR is regarded within the given case study. It was shown, that EV extend the possible power flexibility potential in system states with low power flexibility remarkably and with a high power flexibility also noticeable.

As this paper emphasizes, future grid planning can make use of the FOR as an enhanced method to depict limits within the system operation by varying parameters of the lines and transformers to result in multiple FOR for different constraints and stages of grid expansion. The identification of certain congestion can be obtained when mapping the constraints of all lines and transformers within the examined grid.

The deduced aspects of monetarizing the EV flexibility offer various aspects that need to be analyzed in further research. To point out the applicability of EV as FPU for the provision of ancillary services market designs have to be tested in projects with various participants.

## 8 OUTLOOK

A full utilization of ancillary service potentials at the vertical system interface, e.g. for the ancillary service provision to higher level grid operator, provided by EV or other additional FPU depends on the extension of the grid constraints for some of the investigated scenarios. To map the possibility of using EV with bidirectional charging infrastructure in future flexibility markets from the lowest DSO level to the TSO level, an extensive analysis of real grid topologies has to be carried out. The potential of using the FOR in future grid planning processes can be outlined additionally. This presents a promising approach to gain better insight in expanding grid flexibility potentials.

A furtherly executed scenario analysis extended by time series would offer more detail on the FOR. With focus on the participation in future power flexibility markets in competition to other DER, it has to be examined how costs can be beneficial for EV as market participants. Aspects are the changing power flexibility of, both, EV, which is dependent on the state of charge and the utilized charging strategy, and DEA, which depends on weather conditions and thereby result in varying market prices.

## ACKNOWLODGEMENT

This research with the aim of reduction and removal of grid constraints regarding the integration of charging infrastructure for electric vehicles within the funded project "H-stromert" (01MZ18011B) acknowledges the support of the DLR (Deutsches Zentrum für Luft- und Raumfahrt e. V.).